\documentclass[sigplan,screen]{acmart}
\AtBeginDocument{%
  \providecommand\BibTeX{{%
    \normalfont B\kern-0.5em{\scshape i\kern-0.25em b}\kern-0.8em\TeX}}}

\usepackage{siunitx}[=v2]
\usepackage[htt]{hyphenat}
\usepackage{comment}

\copyrightyear{2023}
\acmYear{2023}
\setcopyright{rightsretained}
\acmConference[EASE '23]{Proceedings of the International Conference on Evaluation and Assessment in Software Engineering}{June 14--16, 2023}{Oulu, Finland}
\acmBooktitle{Proceedings of the International Conference on Evaluation and Assessment in Software Engineering (EASE '23), June 14--16, 2023, Oulu, Finland}\acmDOI{10.1145/3593434.3593472}
\acmISBN{979-8-4007-0044-6/23/06}
\begin{document}

\title{Automating Microservices Test Failure Analysis using Kubernetes Cluster Logs}

\author{Pawan Kumar Sarika}

\affiliation{%
  \institution{Ericsson AB}
  \city{Stockholm}
  \country{Sweden}
}
\email{pawan.sarika@ericsson.com}

\author{Deepika Badampudi}
\affiliation{%
  \institution{Blekinge Institute of Technology}
  \city{Karlskrona}
  \country{Sweden}}
\email{deepika.badampudi@bth.se}

\author{Sai Prashanth Josyula}
\affiliation{%
  \institution{Blekinge Institute of Technology}
  \city{Karlskrona}
  \country{Sweden}}
 \email{sai.prashanth.josyula@bth.se}

\author{Muhammad Usman}
\affiliation{%
   \institution{Blekinge Institute of Technology}
  \city{Karlskrona}
  \country{Sweden}}
  \email{muhammad.usman@bth.se}

\renewcommand{\shortauthors}{Sarika and Badampudi, et al.}

\begin{abstract}
Kubernetes is a free, open-source container orchestration system for deploying and managing Docker containers that host microservices. Kubernetes cluster logs help in determining the reason for the failure. However, as systems become more complex, identifying failure reasons manually becomes more difficult and time-consuming. This study aims to identify effective and efficient classification algorithms to automatically determine the failure reason. We compare five classification algorithms, Support Vector Machines, K-Nearest Neighbors, Random Forest, Gradient Boosting Classifier, and Multilayer Perceptron. Our results indicate that Random Forest produces good accuracy while requiring fewer computational resources than other algorithms.

\end{abstract}



\maketitle

\section{Introduction}
Ericsson, a leading Information and Communication Technology (ICT) service provider, is at the forefront of pioneering cloud RAN\footnote{\url{https://www.ericsson.com/en/ran/cloud}}, a technology that handles network traffic in the cloud. To increase development efficiency, Ericsson developed the Application Development Platform (ADP) ecosystem \cite{usman2022ecosystem}, which among other things, includes a marketplace \textcolor{black}{that hosts over 280 microservices. Among these, 50+ microservices are common microservices} that can be reused and integrated across all applications within the organization. The microservices in the ADP ecosystem are hosted in Docker containers and managed by Kubernetes. 

\textcolor{black}{Some applications need a specific set of microservices that may differ from others}. In addition, the applications may deploy microservices in different environments, such as Amazon Web Service (AWS), or Microsoft Azure Cloud. To ensure high quality, the ADP ecosystem includes a Continuous Integration and Continuous Deployment (CICD) team, which \textcolor{black}{simulates the different application environments} and performs various tests, including testing scalability, robustness, and resilience.

Whenever a test fails, the developers must classify the failure \textcolor{black}{into the following categories - an issue in the cluster, artifactory,  microservice, CICD tests itself, or environment}. The classification is done to report the failure to the relevant team. For example, microservice bugs are assigned to the relevant microservice team, while CICD bugs are reported to the CICD team for resolution. \textcolor{black}{In addition, knowing the failure reason can provide Ericsson insights on what they should improve}. These tests generate an average of 450 MB of log data. Developers use their knowledge and experience to classify the failure, which may take up to two hours for initial analysis. As the number of microservices and teams using them grows, it becomes increasingly challenging and time-consuming for developers to classify these failures manually. 

This paper reports the experience of automating the classification of failures using machine learning techniques at Ericsson. The goal of automation is to identify an efficient and cost-effective method to decrease the time required for manual analysis. To achieve this goal, we utilized developers' knowledge and experiences to better understand the workflows and information necessary for classifying the failure. Based on this understanding, we extracted important log data to train a machine-learning model capable of accurately predicting the failure's cause.

\section{Study Design}

\begin{figure}[h]
  \centering

\includegraphics[width=\linewidth]{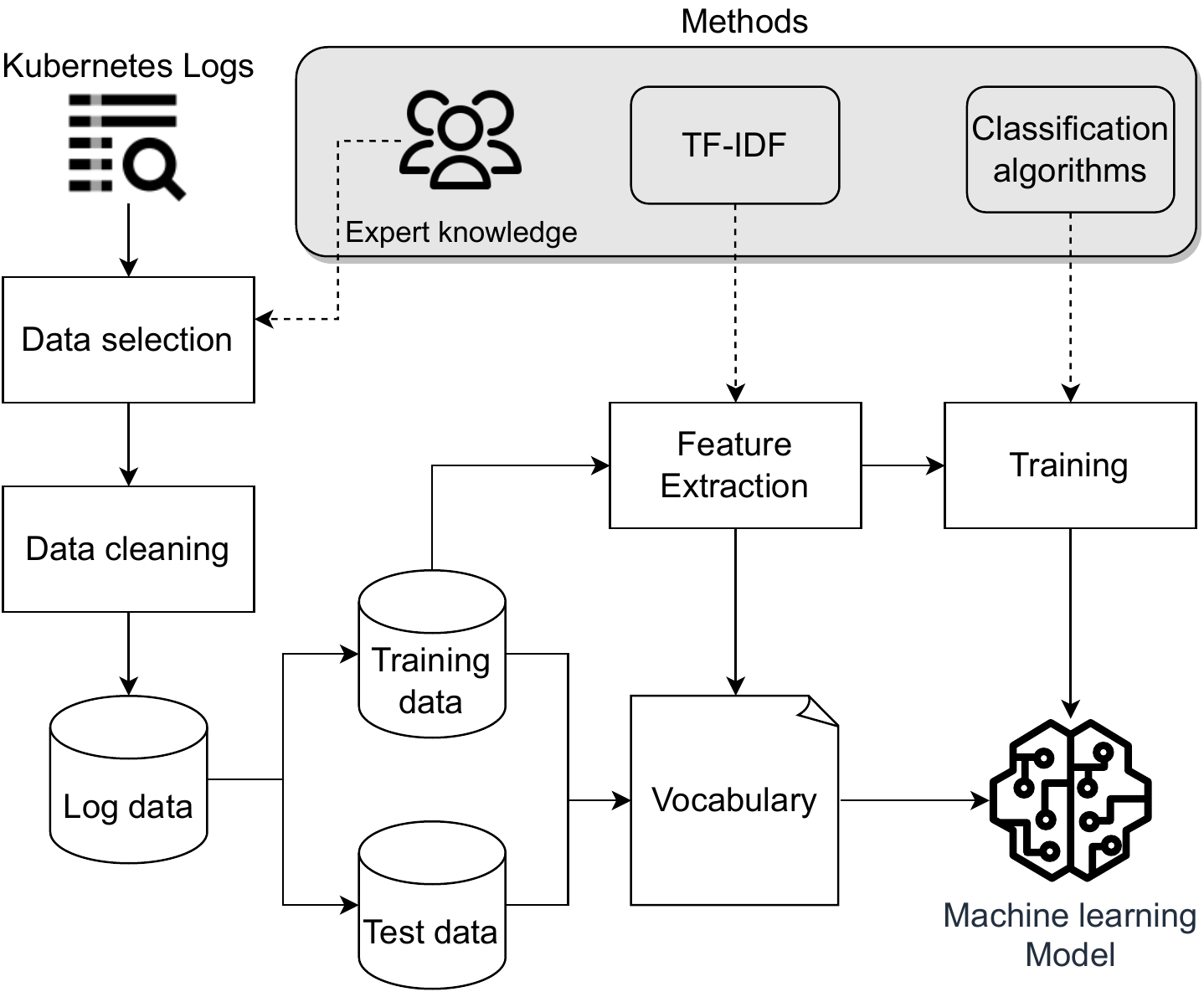}
  \caption{Overall research process.}
  \label{img:experiment_flow_2}
  \Description{Sequence Flow diagram }
\end{figure}

Figure \ref{img:experiment_flow_2} represents the steps followed in this study. In this section, we describe the process followed in each step. 

\subsection{Data Selection}

As depicted in Figure \ref{img:experiment_flow_2}, expert knowledge was utilized to identify the most informative sections of Kubernetes cluster logs, which reduced the overall data processing requirements. Moreover, before automation, it is important to comprehend the manual process of failure classification. The first author, a team member responsible for testing and deploying microservices, noted that each team member had their own approach to analyzing failures. To establish a consensus, we conducted four in-depth interviews to better understand the manual failure classification process. The interviewees had varying levels of experience in identifying reasons for failure, ranging from 1.5 to 4 years. Furthermore, the interviewees have specialized expertise in several areas related to manual failure analysis, such as test environments, end-to-end testing, and coordination with microservice development teams. As a result, the participants have diverse skills and knowledge that were valuable in the failure classification process.


We started the interview by explaining the study objective. We then presented a scenario in which a failure occurred and asked the interviewees to classify the failure using the same steps they would use in a real-world situation. The interviewee is free to use any log as an example. An experienced developer typically examines the sections of the log that are likely to contain the failure reason. This step allowed us to identify the most frequently visited Kubernetes cluster logs.

The interviewees were asked to walk through the different steps they followed, along with the inputs considered and the outputs of each step. Through this process, we identified the crucial log parts for failure classification and how developers identified failure reasons.

\subsection{Data Cleaning}

In this study, we utilized important data identified from the interview for classification. Following data selection, data cleaning was conducted (Figure \ref{img:experiment_flow_2}). We ensured data cleanliness by removing inaccurate, incomplete, or irrelevant data. Human-readable elements (e.g., punctuation, newline characters) were removed to reduce computational costs. We converted all log elements to lowercase letters to avoid ambiguities, and unique elements (e.g., timestamps, IP addresses, line numbers) were removed.  Stopwords and stemming methods were not used as they rely on the English dictionary and may not work for non-English words like 'requesthandlerclass.'


\subsection{Feature Extraction}
After cleaning the Kubernetes cluster logs data, the text format must be converted into a numerical form using Feature Extraction (FE) for algorithms to process it. Although word-to-vector (word2vec) is a popular method for feature extraction, it was not considered appropriate due to the wide variation in the size of each log. Instead, we utilized the TF-IDF method, a common feature extraction procedure for sentences and documents. TF-IDF was more suitable for logs since the words related to the error generally occur once or twice in the entire log. This method highlights words distinctive from the text, allowing the extraction of essential data features required for classification. Using TF-IDF, we generate vocabulary, which is used to extract the feature vectors of the log data. 

\subsection{Classification}
Finally, the most crucial step, i.e., log classification, involves using classification algorithms to construct models that can make accurate predictions. This study employs supervised learning, as it is necessary to make predictions for specified classes. Non-parametric algorithms are selected for this study because they do not require prior knowledge and are particularly suitable when working with large datasets~\citep{j2010artificial}. The chosen algorithms for this study include Support Vector Machines (SVM), K-Nearest Neighbors (KNN), Random Forest Classifier, Gradient Boosting Classifier, and Multilayer Perceptron (MLP).

\subsection{Training}

We proceeded with training the selected classification algorithms using the feature vectors of the training data. We used the \texttt{scikit-learn} library to obtain the trained models. The \texttt{SVC()} is trained with a linear kernel (SVM). The \texttt{RandomForestClassifier()} is trained with 1000 trees. All other function parameters of \texttt{KNeighborsClassifier()} and \texttt{GradientBoostingClassifier()} are set to default values used by \texttt{scikit-learn}. MLP was built using keras sequential model consisting of three dense layers and two Dropout layers. The F1 Score weighted metric is used for evaluation along with the Adam optimizer set to a learning rate of 0.001, and categorical cross-entropy is used as the loss metric.


Following the training of the algorithms, we utilized them to predict class labels for every log in the test data. We collected various metrics, such as accuracy, F1-score, training time, and prediction time, to compare the classification algorithms' performances for predicting failure reasons. We measured training and prediction times to assess computational cost, which can be a better metric for real-world performance than time complexity \cite{lemire_2021}. Stratified 10-fold cross-validation was employed as the sampling strategy. This approach helped identify the algorithm with the best predictive accuracy and computational cost for a given hardware and software configuration.

\section{Expert opinion findings on analyzing failure reasons}

During the interviews, developers scrutinized a failed test case and analyzed the Kubernetes cluster logs to classify the failure. Specifically, they searched for logs associated with the microservice under test within the \texttt{pods/container} or \texttt{pods/describe} directories. If no relevant logs were found, the logs of dependent microservices were consulted. Our analysis revealed that the error or relevant information could usually be found in these directories. Even when it was not, there were clues indicating potential failure type. As a result, we concluded that logs located in \texttt{pods/containers} and \texttt{pods/describe} directories are crucial in most instances. To further strengthen this conclusion, we also examined the tickets submitted by the developers. We found that the root directories for 43 of the 50 reported log snippets were located in \texttt{pods/container} or \texttt{pods/describe}, or both directories.

The \texttt{pods/containers} and \texttt{pods/describe} directories were determined to be important during the interview phase. The logs were pre-processed, and the logs' size was significantly reduced (up to 96.08\%) through data selection and cleaning, reducing computational resources. 

\section{Developing prediction models}
 The classification algorithms were used to process the training data and generate a classification model, which was then used to predict the test data's class labels. The performance of the algorithms was evaluated based on their accuracy, F1-score, training time, and prediction time. 

\begin{table}[!h]
\caption{Model accuracy and F1-score for each algorithm}
\label{tab:accuracy-f1}
\begin{tabular}{lcc}
\toprule
Algorithm & Accuracy & F1-score \\
\midrule
SVM & 0.6905 & 0.6580 \\
KNN & 0.6576 & 0.6580 \\
Random Forest & 0.7279 & 0.7106 \\
Gradient Boosting & 0.7302 & 0.7071 \\
MLP & 0.7086 & 0.6911 \\
\bottomrule
\end{tabular}
\end{table}


We compared the chosen classification algorithms statistically based on the recommendations by \citet{demvsar2006statistical}. We applied the Friedman test to test the null-hypothesis that all algorithms are equivalent. If this hypothesis was rejected, we conducted a post-hoc Nemenyi test to compare the algorithms pairwise. 

Table~\ref{tab:accuracy-f1} presents the accuracy and F1-scores of each algorithm. The Friedman test was performed to evaluate the statistical significance of the differences between the algorithms, resulting in a p-value of 0.00074 and a Q value of 19.12, confirming significant differences in accuracy. We subsequently conducted a Nemenyi test, and the p-values of the combinations of algorithms are shown in Table~\ref{tab:p-value_accu_exp1} with P-values less than $\alpha$ (0.05) in bold. Random Forest and Gradient Boosting were statistically better than KNN.



\begin{table}[]
\centering
\caption{Pairwise P-values for accuracy}
\label{tab:p-value_accu_exp1}
\resizebox{0.45\textwidth}{!}{%
\begin{tabular}{|l|l|l|l|l|l|}
\hline
 &
  \textbf{SVM} &
  \textbf{KNN} &
  \textbf{\begin{tabular}[c]{@{}l@{}}Random \\ Forest\end{tabular}} &
  \textbf{\begin{tabular}[c]{@{}l@{}}Gradient \\ Boosting\end{tabular}} &
  \textbf{MLP} \\[5pt] 
  \hline 
\textbf{SVM}                                                         & 1.0000 & 0.3517 & 0.4374          & 0.2418          & 0.9000 
\\\hline 
\textbf{KNN}                                                         &       & 1.0000 & \textbf{0.0037} & \textbf{0.0010} & 0.1143 
\\\hline  
\textbf{\begin{tabular}[c]{@{}l@{}}Random
\\ Forest\end{tabular}}    &       &          & 1.0000        & 0.9000          & 0.7633 
\\\hline  
\textbf{\begin{tabular}[c]{@{}l@{}}Gradient\\ Boosting\end{tabular}} &       &          &               & 1.0000          & 0.5626
\\\hline 
\textbf{MLP}                                                         &       &          &               &           & 1.0000 \\
\hline
\end{tabular}%
}
\end{table}

Similarly, for F1-scores (Table~\ref{tab:p-value_f1_exp1}), the Friedman test yielded a p-value of 0.00056 and a Q value of 19.76, indicating significant differences. The Nemenyi test showed that Random Forest outperformed SVM and KNN with statistically significant differences. Furthermore, KNN's performance was significantly lower than Gradient Boosting's. Regarding both F1-score and accuracy, Random Forest, Gradient Boosting, and MLP performed similarly.

We perform the Nemenyi test to determine which algorithms are statistically different based on F1-scores. Table \ref{tab:p-value_accu_exp1} represents the p-value of F1-score for each combination of algorithms. The p-value in bold is less than $\alpha$ (0.05). 
\begin{table}[!ht]
\centering
\caption{Pairwise P-values for F1-score}
\label{tab:p-value_f1_exp1}
\resizebox{0.45\textwidth}{!}{%
\begin{tabular}{|l|l|l|l|l|l|}
\hline
 & SVM & KNN & \begin{tabular}[c]{@{}l@{}}Random \\ Forest\end{tabular} & \begin{tabular}[c]{@{}l@{}}Gradient \\ Boosting\end{tabular} & MLP \\ \hline
SVM                                                         & 1.0000 & 0.9000 & \textbf{0.0248} & 0.1143 & 0.4374 \\ \hline
KNN                                                         &  & 1.0000 & \textbf{0.0022} & \textbf{0.0160} & 0.1143 \\ \hline
\begin{tabular}[c]{@{}l@{}}Random \\ Forest\end{tabular}    &  &  & 1.0000 & 0.9000 & 0.6830 \\ \hline
\begin{tabular}[c]{@{}l@{}}Gradient\\ Boosting\end{tabular} &  &  &  & 1.0000 & 0.9000 \\ \hline
MLP                                                         &  &  &  &  & 1.0000 \\ \hline
\end{tabular}%
}
\end{table}



Finally, the algorithms were evaluated based on their training and prediction times. The average training time in minutes for all ten folds is shown in Table \ref{tab:time}. It was observed that Gradient Boosting took a significantly longer time for training when compared to other algorithms. On the other hand, MLP took more than 2x longer for training compared to the Random Forest. Regarding prediction time, all five algorithms finished predictions in less than a second, indicating no significant difference between them in real-world performance. Based on the analysis, it can be concluded that the Random Forest algorithm performed better in terms of both accuracy and training time.

\begin{table}[!h]
  \caption{Training and prediction time in minutes}
  \label{tab:time}
  \begin{tabular}{lS[table-format=3.2]S[table-format=3.2]}
    \toprule
     Algorithm & \text{Training time} & \text{Prediction time} \\
    \midrule
    SVM & 0.845 & 0.082 \\
KNN & 0.004 & 0.015 \\
Random Forest & 1.702 & 0.006 \\
Gradient Boosting & 110.773 & 0.002 \\
MLP & 4.111 & 0.005 \\
    \bottomrule
  \end{tabular}
\end{table}

\section{Piloting the prediction models}

To test the model in a production environment, we developed an end-to-end prototype. The results of the model were added to a dashboard for developer analysis. 

 We received feedback from developers, who indicated that while the failure classification helps them understand the reason for the failure and where they should look in the logs, they would appreciate additional information beyond the failure reason. As part of future work, we will devise an additional solution to include the most similar log previously reported along with its diagnosis report. Reviewing diagnosis reports of similar logs will help developers quickly understand the cause of failure. This way, developers can precisely look at the specific logs increasing their efficiency and helping them close the cases faster.    

In addition, developers expressed concerns about the trustworthiness of the models' predictions due to a lack of understanding of how the model arrived at its conclusions. Further effort is required to improve the models' interpretability and explainability. 

Although Random Forest performed better, we need to tune the hyperparameters to truly capture the effectiveness of Random Forest compared to other algorithms.

\section{Conclusion}
Identifying the reason for a failed CICD test is important to assign the failure to the relevant team for resolution. The automatic identification of failure reasons minimizes the manual effort needed to analyze the failure. In this study, we report experiences of classifying large-scale Kubernetes cluster logs using machine learning classification algorithms: Support Vector Machines (SVM), K-Nearest Neighbors (KNN), Random Forest Classifier, Gradient Boosting Classifier, and Multilayer Perceptron (MLP). Each log generates 450 MB of data on average, which was reduced to 91\% by identifying relevant parts of the logs using expert opinion. We used the TF-IDF feature extraction method. Then we trained and evaluated the classification algorithms in terms of accuracy, F1 scores, training, and prediction time.

The results indicated that Random Forest outperforms SVM, KNN, Gradient Boosting, and MLP in terms of accuracy and computational cost. When evaluating the proof-of-concept, the developers expect additional information to strengthen the confidence in the prediction. For example, providing similar logs and diagnosis reports can help the developers to understand how previously failed tests were classified manually and arrived at the same classification predicted by the model. As part of future work, we aim to make the comparison of the algorithms more rigorous by applying hyperparameter tuning and further improving the model's interpretability and explainability. 

As a prospect for future research, we intend to enhance the dimensionality of the input data. Currently, we concatenate logs from different services into a single file, creating a two-dimensional input. Our plan is to aggregate logs from different pods within each service, producing a three-dimensional input where each service has its own file. This approach would enable us to leverage deep learning algorithms, such as CNNs, which are proficient in processing 3D data. By doing so, our model would be able to identify the interdependencies among the services, leading to improved results.
 
\begin{acks}
The Knowledge Foundation supports this work through the OSIR project (reference number 20190081) at Blekinge Institute of Technology, Sweden.
\end{acks}
\bibliographystyle{ACM-Reference-Format}
\bibliography{sample-base}


\end{document}